\documentclass[conference]{IEEEtran}
\IEEEoverridecommandlockouts
\usepackage{cite}
\usepackage{amsmath,amssymb,amsfonts}
\usepackage{algorithmic}
\usepackage{graphicx}
\usepackage{textcomp}
\usepackage{rotating}
\usepackage{array}
\usepackage{xcolor}
\usepackage{tikz}
\usepackage{makecell}
\usepackage{multirow}
\usepackage{etoolbox}
\def\BibTeX{{\rm B\kern-.05em{\sc i\kern-.025em b}\kern-.08em
    T\kern-.1667em\lower.7ex\hbox{E}\kern-.125emX}}
\begin{document}

\newcommand*\circled[1]{\tikz[baseline=(char.base)]{
		\node[shape=circle,draw,inner sep=0.8pt] (char) {#1};}}

\title{Project-Level Encoding for Neural Source Code Summarization of Subroutines
	\thanks{This work is supported in part by NSF CCF-1452959 and CCF-1717607 grants.
}}

\author{\IEEEauthorblockN{Aakash Bansal, Sakib Haque, and Collin McMillan}
\IEEEauthorblockA{\textit{Dept. of Computer Science and Engineering} \\
	\textit{University of Notre Dame}\\
	Notre Dame, IN, USA \\
	\{abansal1, shaque, cmc\}@nd.edu
}}

\maketitle

\begin{abstract}
Source code summarization of a subroutine is the task of writing a short, natural language description of that subroutine.  The description usually serves in documentation aimed at programmers, where even brief phrase (e.g. ``compresses data to a zip file'') can help readers rapidly comprehend what a subroutine does without resorting to reading the code itself.  Techniques based on neural networks (and encoder-decoder model designs in particular) have established themselves as the state-of-the-art.  Yet a problem widely recognized with these models is that they assume the information needed to create a summary is present within the code being summarized itself -- an assumption which is at odds with program comprehension literature.  Thus a current research frontier lies in the question of encoding source code context into neural models of summarization.  In this paper, we present a project-level encoder to improve models of code summarization.  By project-level, we mean that we create a vectorized representation of selected code files in a software project, and use that representation to augment the encoder of state-of-the-art neural code summarization techniques.  We demonstrate how our encoder improves several existing models, and provide guidelines for maximizing improvement while controlling time and resource costs in model size.
\end{abstract}


\section{Introduction}

Source code summarization is the task of writing short, natural language descriptions of that code~\cite{haiduc2010use, sridhara2010towards}.  Typical targets of summarization are the subroutines of a software project.  The purpose of the descriptions is to provide human readers with a big picture view of what each subroutine does.  Even a single phrase e.g. ``compresses data to a zip file'' can help a person understand code without having to read every detail of that code~\cite{forward2002relevance}.  Summaries of subroutines form the foundation of much documentation aimed at programmers such as JavaDocs~\cite{kramer1999api}, and the literature is replete with studies demonstrating how programmers often rely on these summaries, only turning to reading the code itself as a last resort~\cite{xia2017measuring}.  And while a majority of documentation is still written manually, recent research has made inroads towards automatic code summarization~\cite{zhao2020survey}.


The backbone of almost all state-of-the-art approaches to automatic source code summarization is the neural encoder-decoder model architecture~\cite{leclair2019neural, zugner2021languageagnostic, hu2018deep}.  This architecture has its roots in machine translation~\cite{bahdanau2014neural}, in which the encoder creates a vectorized representation of a sentence in one language (e.g. English), while the decoder creates a representation of that same sentence in a different language (e.g. French).  When trained with enough data (on the order of millions of examples~\cite{leclair2019recommendations}), these models can learn to associate patterns in the encoder representation to patterns in the decoder representation.  After training, the encoder can be given an input example, and the model can generate a likely decoder representation and therefore a likely output example -- and translate French sentences to English.  For source code, the encoder's job is to represent the source code, while the decoder represents the source code summary -- give the encoder source code, and the decoder generates a summary.

The obvious problem with these approaches is that they can only generate a summary based on whatever source code is passed to the encoder.  Thus these approaches make a tacit assumption that all of the information necessary to generate that summary is present in that source code.  This assumption is at odds with decades of program comprehension literature~\cite{Siegmund:2014:UUS:2568225.2568252, letovsky1987cognitive, von1995program}.  This literature is quite clear that high-level descriptions such as summaries very often contain concepts that can only be understood in the context of the other code in the same software project.  To paraphrase a classic example, a subroutine called {\small \texttt{book()}} can only be fully understood if it is also known that it exists in a class called {\small \texttt{Seats}} in a project called {\small \texttt{AircraftTravel}}~\cite{biggerstaff1993concept}.

In this paper, we present a project-level encoder to augment existing encoder-decoder neural models of source code summarization.  Our approach is ``project level'' in that it creates a vectorized representation of a subset of code files in a software project.  Our approach augments existing models in that the output of our approach may be combined with encoder portion of most existing code summarization models: most models contain an encoder for the source code itself that produces some vectorized representation of that code, and our encoder extends that representation.  The advantage to our encoder is that it provides context to the model about the software project in which a subroutine exists, so that the model does not rely only on the information in that subroutine.

We evaluate our project encoder in three ways.  First, we implement our project encoder as an addition to four existing neural source code summarization techniques.  We demonstrate that our encoder boosts the performance of these techniques by between 4 and 8\% in terms of BLEU scores in a large Java dataset, and between 9\% and 17.5\% in ensemble models in that dataset.  Second, we compare our whole-project encoder with a competitive approach that attempts to summarize the context surrounding code, and found between 1.5\% and 7\% improvement in terms of BLEU score in the Java dataset.  Third, we study the time and resource costs of our project encoder, to determine the costs associated with the increased performance of our approach.

We provide all data and implementations via our online appendix (see Section~\ref{sec:repro}).

\section{Background \& Related Work}

This section describes related work and supporting technologies, namely the neural encoder-decoder architecture.

\begin{figure}[!b]
	\centering
	{\small
		\vspace{-0.4cm}
		\begin{tabular}{p{3.9cm}p{0.4cm}p{0.4cm}p{0.4cm}p{0.4cm}p{0.4cm}}
			& E          & S          & C                \\
			\textcolor{white}{*}Loyola~\emph{et al.}~(2017)~\cite{loyola2017neural}					& x &  	&      \\
			\textcolor{white}{*}Lu~\emph{et al.}~(2017)~\cite{lu2017learning}						& x &  	&      \\
			\textcolor{white}{*}Jiang~\emph{et al.}~(2017)~\cite{jiang2017automatically}			& x &  	&      \\
			\textcolor{white}{*}Hu~\emph{et al.}~(2018)~\cite{hu2018summarizing}					& x &  	&      \\
			\textcolor{white}{*}Hu~\emph{et al.}~(2018)~\cite{hu2018deep}							& x & x	&      \\
			\textcolor{white}{*}Allamanis~\emph{et al.}~(2018)~\cite{allamanis2018learning}			& x & x	&      \\
			\textcolor{white}{*}Wan~\emph{et al.}~(2018)~\cite{wan2018improving}					& x & x	&      \\
			\textcolor{white}{*}Liang~\emph{et al.}~(2018)~\cite{liang2018automatic}				& x & x	&      \\
			\textcolor{white}{*}Alon~\emph{et al.}~(2019)~\cite{alon2019code2seq, alon2019code2vec}	& x & x	&      \\
			\textcolor{white}{*}Gao~\emph{et al.}~(2019)~\cite{gao2019neural}						& x &  	&      \\
			\textcolor{white}{*}LeClair~\emph{et al.}~(2019)~\cite{leclair2019neural}				& x & x	&      \\
			\textcolor{white}{*}Mesbah~\emph{et al.}~(2019)~\cite{mesbah2019deepdelta}				& x & x	&      \\
			\textcolor{white}{*}Nie~\emph{et al.}~(2019)~\cite{nie2019framework}					& x & x	&      \\
			\textcolor{white}{*}Haldar~\emph{et al.}~(2020)~\cite{haldar2020multi}					& x & x	&      \\
			\textcolor{white}{*}Ahmad~\emph{et al.}~(2020)~\cite{ahmad2020transformer}				& x &  	&      \\
			\textcolor{white}{*}Haque~\emph{et al.}~(2020)~\cite{haque2020improved}					& x &  	& x    \\
			\textcolor{white}{*}Z{\"u}gner~\emph{et al.}~(2021)~\cite{zugner2021languageagnostic}	& x & x &      \\
			\textcolor{white}{*}Liu~\emph{et al.}~(2021)~\cite{liu2021retrievalaugmented}			& x & x &      \\
			\textcolor{white}{*}\emph{$<$this paper$>$}													& x &   & x
		\end{tabular}
	}
	\vspace{0.1cm}
	\caption{\small{Key peer-reviewed related work from the last four years.  Column $E$ means the approach is an encoder-decoder architecture.  $S$ means that the improvement of the model relies primarily on structural information about the source code being summarized, such as a subroutine's AST.  $C$ means the improvement is primarily due to contextual information.}}
	\label{tab:screlated}
\end{figure}

\subsection{Source Code Summarization}
\label{sec:related_scs}

Figure~\ref{tab:screlated} shows key papers related to source code summarization in the last four years.  The list is not exhaustive and only includes peer-reviewed work.  Papers are broadly categorized as based on the encoder-decoder architecture (column $E$), and whether their novelty (and primary means of improvement over baselines) is based on structural information about the source code itself (column $S$), or contextual information about surrounding source code (column $C$).  Two observations are apparent.  First, recent approaches are based on some variant of an encoder-decoder architecture.  Prior to 2017, code summarization research focused on templates or information retrieval, but these have recently given way to neural encoder-decoder designs~\cite{zhao2020survey, mcburney2014automatic, iyer2016summarizing, sridhara2010towards}.

A second observation is that the strong trend has been to squeeze ever more information out of the source code being summarized itself, in the form of structural information.  The trend began around the time Hu~\emph{et al.}~\cite{hu2018deep} used the abstract syntax tree to mark up the source code tokens in the encoder's input sequence, and was followed up by Allamanis~\emph{et al.}~\cite{allamanis2018learning}, LeClair~\emph{et al.}~\cite{leclair2019neural}, among others noted in the table, with different AST-based code representations.  Advancement continued as AST path-based encoders~\cite{alon2018code2seq, alon2019code2vec} were followed by AST graph neural network-based encoders~\cite{leclair2020improved}.  While some research has been dedicated to novel representations of the text in code (e.g. via Transformer models~\cite{ahmad2020transformer}), the tendency has been towards more and more complex representations of the code structure.  Very recently, multi-edge and hybrid GNN structures have been devised~\cite{zugner2021languageagnostic, liu2021retrievalaugmented}.

Much more rare is work that attempts to improve performance by integrating contextual information.  Code context may be broadly defined as the source code in the methods, files, and packages surrounding a particular snippet of code~\cite{krinke2006effects}.  Program comprehension literature is quite clear that the context surrounding source code is critical to understanding that code, with work ranging from psychological/physiological studies~\cite{Siegmund:2014:UUS:2568225.2568252, letovsky1987cognitive, von1995program} to empirical/technical solutions~\cite{maalej2014comprehension, rajlich2002role, roehm2012professional, maletic2001supporting} verifying this conclusion.  The use of code context for summarization was mainstream among older template- and IR-based techniques~\cite{neubigsurvey, mcburney2014automatic, sridhara2010towards}, though it is currently overlooked among neural network-based solutions.  Haque~\emph{et al.}~\cite{haque2020improved} are a notable exception.  They use text from each function in the same file as part of the encoder portion of their model, and show improvement over different baselines.

This paper focuses on contextual information.  Specifically, we focus on project context, which is the context provided by every source code file in the same project as the subroutine we are summarizing.  This context is more broad than the file-level context proposed by Haque~\emph{et al.}~\cite{haque2020improved}, but like that work, our approach is complementary to most encoder-decoder approaches rather than competitive.  Our approach augments the solutions based on the structure of the code itself, it does not replace them.

\subsection{Neural Encoder-Decoder Architecture}

The neural encoder-decoder architecture revolves around two independent vectorized representations of parallel data.  The parallel data may be a sentence in one language and its translation in another language, an image and a caption of that image, or a subroutine and a natural language summary of that subroutine.  Since each ``side'' of the parallel data may be quite different, the means of generating the vectorized representation will also be different.  Usually the purpose of an encoder-decoder model is to create one ``side'' of the data out of the other (e.g. create a summary out of a subroutine).  The input side is referred to as the encoder, and the output side is referred to as the decoder.  Thus to train a model to e.g. translate from French to English, the encoder receives French sentences and the decoder receives the parallel English sentences.  The encoder-decoder architecture has its roots in work by Sutskever~\emph{et al.}~\cite{sutskever2014sequence} published 2014.  Since then the architecture has blossomed and found an extremely wide variety of uses, as several survey papers testify~\cite{young2018recent, pouyanfar2018survey, shrestha2019review}.

The vast majority of encoder-decoder architectures work because of a similarity calculation that links the encoder and decoder representation, called an attention mechanism.  Essentially what the attention mechanism does is compute the similarity between the encoder and decoder representations, which helps the model learn to associate features in those representations.  For example, a single word in a French sentence would be associated with its counterpart in the English sentence.  Attention was proposed by Bahdanau~\emph{et al.}~\cite{bahdanau2014neural} and has become an integral part of most encoder-decoder models.

This paper follows in the tradition of most encoder-decoder models, though with a small twist to the attention mechanism.  As the next section will show, we maintain an independent attention mechanism for our whole-project encoder, so the model can learn to attend to both the encoder for that subroutine and our whole-project encoder.  In effect, the model will learn from both a local context of the subroutine itself and a global context of the whole software project.  When defined in these terms, our work is related to ``cascade attention'' from image processing~\cite{zhu2020learning, wang2017cascade, sun2018context, li2020ensemble}.  For example, work by Wang~\emph{et al.}~\cite{wang2018cascade} detects human emotion with a closeup of a person's face and also a zoomed out image of the entire room.  The ``cascade'' is that the model attends to both the closeup and the zoomed out image.  The idea is that it may detect crying in a face, but then understand it as either sadness or happiness depending on the context.  Likewise, our approach is to learn from the subroutine's source code (via any number of existing encoders), then form a better understanding of it with our whole-project encoder.

\section{Our Approach}

Our approach, in a nutshell, is to create an encoder of a selection of the files in the same project as a subroutine, then combine this encoder with an arbitrary encoder of the subroutine itself.  This section starts with our definition of project context.  We then provide an overview of the encoder and guidelines for combining with existing models.

\subsection{Project Context}

We define ``project context'' as all source code files of the same language in the whole software project in which a subroutine resides.  For example, for a Java method, the project context would be all other Java files in the same project.  The advantage of this broad definition of project context is that it allows the model to learn from the high level concepts that are described in many areas of the project (we illustrate this advantage with examples in Section~\ref{sec:descr}).  A potential disadvantage is that the project context will often be very large.  A risk is that the model size could become so large that it is not feasible due to time or resource constraints -- at the time of writing, not every user may be expected to have a GPU with 16gb VRAM, for instance.  While we study this risk in RQ$_4$, controlling these potential costs is a key factor in our model design.  Therefore, even though project context is defined as all source code files, not all information from all files will be included in the model.

\subsection{Model Overview}
\label{sec:modeloverview}

Our model centers around four vector spaces: the word, subroutine, file, and project embeddings.  The input to these is regulated by five hyperparameters, noted in the figures below:

\begin{figure}[h!]
	\vspace{-0.3cm}
	\centering
	\includegraphics[width=0.35\textwidth]{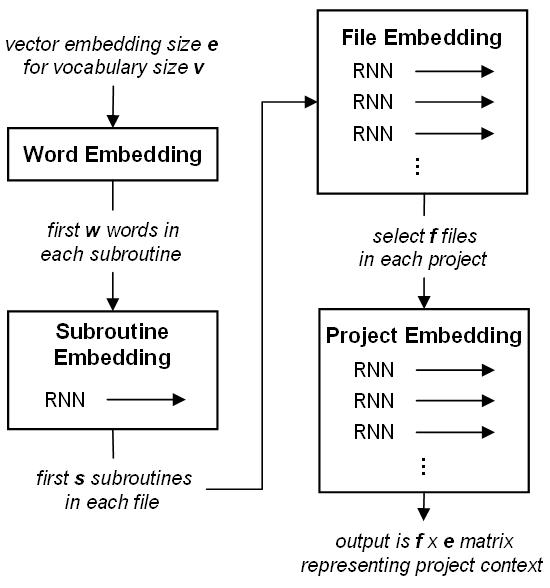}
	\vspace{-0.3cm}
\end{figure}

\emph{Word Embedding} The word embedding is identical to that presented in many papers on neural NLP topics, including code summarization.  Essentially, each word is represented as an $e$-length vector.  To control model size, a maximum of the first $v$ most-common words is included in the embedding space, others are marked with a default out-of-vocab token, also in the embedding space.  A ``word'' in source code is defined by the preprocessing procedure.  We use the preprocessor recommended for code summarization by LeClair~\emph{et al.}~\cite{leclair2019recommendations}.

\emph{Subroutine Embedding} The subroutine embedding results in a vectorized representation of each subroutine.  The input to the subroutine embedding is the word embedding vector for the first $w$ words in the subroutine.  Then we pass these words as a sequence through a recurrent neural network.  The final state of that RNN is the vectorized representation of the subroutine.  We chose to use the first $w$ words (as opposed to $w$ random words, words ranked via tf/idf, etc.) because these words will include the signature of the method, which has been shown to be the most important component for summarization~\cite{rodeghero2014improving}.

\emph{File Embedding} The file embedding results in a vectorized representation of each file.  The input is the embedding for the first $s$ subroutines in a file -- an $s$ x $e$ matrix because each subroutine is represented with an $e$-length vector.  We then use each row in the matrix as a position in a sequence, which we send to an RNN.  The final state of the RNN is the file embedding vector.  An RNN is a reasonable choice to combine subroutine vectors because the subroutines occur in the file in an order defined by the author of that file.  The meaning of this order may be disputed, however, so future work may consider aggregating these vectors by some other means such as averaging.

\emph{Project Embedding} The project embedding output is the final output of the project encoder, prior to applying attention.  The input is the file embedding for $f$ files in the project.  We chose these files from the project with an operation {\small \texttt{SELECT}}.  In our implementation, the {\small \texttt{SELECT}} operation randomly chooses $f$ files from the project for each subroutine -- each subroutine has a new $f$ random selections.  The output of the project embedding is an $f$ x $e$ matrix in which each row is a file embedding and each column is an index in the vector representation of those files.  Note that we do not aggregate this matrix into a single vector.  The reason is that we use an attention mechanism (not shown in the figure above) to attend each position in the decoder to each position in this project embedding.  The design of our attention mechanism is identical to Luong~\emph{et al.}~\cite{luong2015effective}, though in principle another may be used.  The result is that the model will learn to attend to the most important files in the project embedding.  We present an example of how attention to the project embedding helps the model in Section~\ref{sec:descr}.

\subsection{Implementation Guidelines}

Our implementation guidelines fall into two categories: hyperparameter/setup recommendations, and suggestions for integration with other encoder-decoder models.

\subsubsection{Hyperparameters}

While a grid search for every parameter is not feasible due to high computation costs, we chose the following based on both related literature and pilot tests:

\vspace{-0.175cm}
\begin{table}[h!]
	\centering
	\begin{tabular}{lll}
		$e$   & 100   & vector length \\
		$v$   & 10000 & vocab size \\
		$w$   & 25    & words per subroutine \\
		$s$   & 10    & subroutines per file \\
		$f$   & 10    & files per project \\
		$RNN$ & GRU   & type of RNN
	\end{tabular}
\end{table}
\vspace{-0.225cm}

The values of 100 for $e$ and 10,000 for $v$ are based on successful results and recommendations by LeClair~\emph{et al.}~\cite{leclair2019recommendations} for neural source code summarization.  The value for $w$ is based on findings that the signature of a subroutine typically contains the most valuable textual information about that subroutine (since it neatly condenses the return type, name, and parameter types in a few words) -- we chose 25 because that value covers the entire signature in a majority of subroutines and because several RNN designs have been shown to lose the ability to preserve dependencies when the sequence becomes too long.  We chose a GRU as the RNN as a balance between ability to preserve dependencies and time cost of computation.

The values for $s$ and $f$ are more subjective.  On the one hand, maximizing these numbers means the model can consider much more of the project context.  But on the other hand, computation and memory cost will rapidly become prohibitive.  Consider that one file is 100kb of memory in the model ($s$ x $w$ x $e$ x 4 bytes = 10 x 25 x 100 x 4 bytes).  Each project context matrix is then 1mb ($f$ = 10 files).  The costs add up because the datasets may involve millions of subroutines.

\subsubsection{Integration}
\label{sec:impl}

Recall that our intent for our whole-project encoder is to be integrated with the encoder portion of an existing encoder-decoder model.  The simplest means of integration is to treat the whole-project encoder as independent of all other parts of the model, and connect its output to the existing encoder's output after attention is applied.  Consider a ``vanilla'' seq2seq-like model like those used in the first neural code summarization papers~\cite{iyer2016summarizing, hu2018deep}, that has a single RNN as an encoder of the words in the subroutines and a single RNN as the decoder for the words in the summary.  This model would typically have an attention mechanism between the encoder and decoder which would adjust the emphasis of the information in the encoder based on the decoder.  Then the attended encoder output would be combined to the decoder output and connected to a fully-connected output, as below:

\begin{figure}[h!]
	\vspace{-0.4cm}
	\centering
	\includegraphics[width=0.35\textwidth]{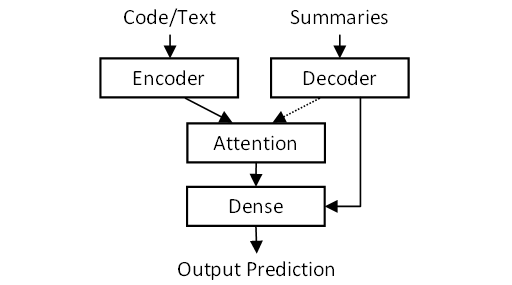}
	\vspace{-0.25cm}
\end{figure}

Integrating this model with our whole-project encoder would involve rewiring the output of the decoder to an attention mechanism for the whole-project encoder (in addition to the pre-existing encoder), as mentioned under the Project Embedding heading in the previous section.  Then the output of the existing encoder, the whole-project encoder, and the decoder would be combined and connected to the fully-connected output layer:

\begin{figure}[h!]
	\vspace{-0.3cm}
	\centering
	\includegraphics[width=0.35\textwidth]{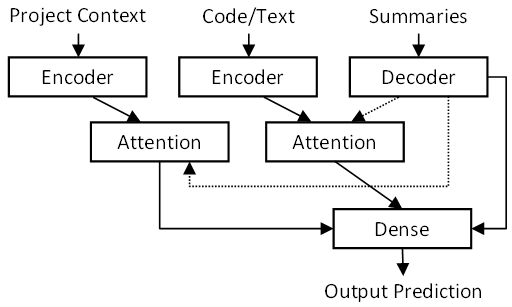}
	\vspace{-0.25cm}
\end{figure}

Three key details stand out as questions for this implementation (for maximum clarity, readers may elect to follow along in our implementation in file {\small\texttt{models/attendgru\_pc.py}} in our online appendix (Section~\ref{sec:repro})).  First, we combine the output of the encoders and the decoder by concatenating the matrices, which triples the vector length.  We then use a fully-connected layer to squash these long vectors back to the specified vector size $e$ (lines 80-85 in the implementation).  The effect is that the model learns how to combine the vectors during training.  

Second, we share the word embedding space between the code/text encoder and the whole-project encoder (around line 64 in the implementation).  This is possible because both models use the same vocabulary, and saves both memory space and computation time.  Separate word embeddings would be necessary for different vocabularies, or if it is desirable to use a pretrained embedding, etc.

Third, the project context input is separate from the other encoder/decoder input, and must be extracted from the dataset prior to training.  This requirement is not likely to be a problem for models that already use the source code of the subroutines to generate summaries, but there may exist application domains where this data is not available.




\section{Evaluation}

In this section, we describe our evaluation, including research questions, methodology, datasets, and baselines.

\subsection{Research Questions}

Our research objective is to determine the degree of difference in performance that our whole-project encoder imbues on other recent neural code summarization techniques.  We explore this difference from several angles by asking the following Research Questions (RQs):

\begin{description}
	\item[RQ$_{1}$] What is the difference in performance of recent baselines when augmented with our encoder, according to standard quality metrics over a large dataset?
	
	\vspace{0.05cm}
	
	\item[RQ$_{2}$] What is the difference in performance when measuring only the action word prediction quality?
	
	\vspace{0.05cm}

	\item[RQ$_{3}$] What is the difference in performance compared to a baseline code context encoder?
	
	\vspace{0.05cm}

	\item[RQ$_{4}$] What is the cost of the performance increase in terms of model time to train?
	


\end{description}

The rationale behind RQ$_1$ is that the vast majority of neural code summarization research uses a set of established metrics (namely, BLEU) to evaluate the quality of the predicted summaries to a gold set.  We follow this practice to compare the baselines to our augmented versions of those baselines.   Evaluations in many papers stop here.  Yet, complaints are rising that the accepted evaluation practice may not capture quality to the extent desired~\cite{stapleton2020human}, so we ask several more RQs to give a more complete picture.

We ask RQ$_2$ in light of new recommendations by Haque~\emph{et al.}~\cite{haque2021action} on evaluating code summarization techniques.  They observe that an overwhelming majority of code summaries start with an action word, in keeping with style guide recommendations (e.g. ``connects to game server'' or ``adds row to sql table'').  They find that this action word is a critical piece of the prediction quality, and recommend that the quality of the prediction of these words should be assessed independently from the assessment of the entire summary output.

The purpose of RQ$_3$ is to evaluate our approach against a baseline for contextual information.  Recall that most code summarization techniques focus on the code within a subroutine (or other code snippet) itself.  Our approach augments these techniques.  However, at least one other code context encoder has been proposed, so we evaluate against it.

The rationale behind RQ$_4$ is that adding our whole-project encoder will impose some time cost over the baselines, and recent work from industry reports that training time due to added model complexity creates engineering difficulties in practice~\cite{hazelwood2018applied}. We study this cost to guide cost-benefit analysis to using our approach.


\subsection{Methodology}
\label{sec:metho}

Our methodology for answering RQ$_1$ closely follows the procedures of almost all neural source code summarization papers to date.  We obtain two datasets of subroutines and summaries of those subroutines, and divide them into training/validation/test subsets (our dataset preparation procedures described in the next section).  Then we train each of our baselines (also described below) with these datasets to a maximum of 10 epochs.  We use the teacher forcing training procedure, as do most recent code summarization papers~\cite{alon2018code2seq, leclair2020improved}.  We chose the trained model from the epoch that achieved the highest validation accuracy, so each baseline has the opportunity to find an optimum within a reasonable training time ceiling (each epoch for most models takes 2-3 hours, so ten epochs is approximately 24 hours).  Then we use each baseline's trained model to predict a code summary for the subroutines in the test set.  Finally, we report the BLEU~\cite{Papineni:2002:BMA:1073083.1073135} and ROUGE-LCS~\cite{lin2004rouge} scores for each baseline.  We repeat the entire procedure for our versions of the baselines that we augment with the whole-project encoder.  Note that this procedure for RQ$_1$ is not novel -- our intent is to adhere to community standards.

We elected to focus on an in-depth metrics-driven evaluation rather than a human study.  While human studies are often considered a gold standard for evaluation, Chatzikoumi~\emph{et al.}~\cite{chatzikoumi2020evaluate} point out that reality is more nuanced.  Human studies are very valuable, but have two key problems.  First, they are not reproducible because people are subject to biases, fatigue, mistakes, and other factors, so people may give very different results.  Second, humans can only be expected to evaluate a few dozen or hundred samples.  In this paper, we have 24 model configurations to test, and a test set with tens of thousands of samples.  Therefore, we decided to focus on an in-depth analysis of metrics-driven evaluation.


To answer RQ$_2$, we follow the recommendations of Haque~\emph{et al.}~\cite{haque2021action}.  The training process is almost identical to RQ$_1$.  The difference is that a filter based on the Stanford NLP package~\cite{manning2014stanford} is used to extract the action word from the gold set summary (in practice, this is usually the first word), and the model is trained to predict just that word.  Then during testing, the model is asked to predict the action word for the subroutines in the test set.  Precision and Recall~\cite{tharwat2020classification} are used to assess the quality of the predictions for each action word: precision is the percent of predictions of that action word which were correct, recall is the percent of instances of that action word in the gold set that are predicted.  The macro average of these precision and recall values across all action words is reported.  For clarity, we also produce confusion matrices to the extent possible within page limits.

For brevity, we select only a subset of the best-performing approaches from RQ$_1$ and RQ$_2$ as representative examples for the remaining RQs.  However, we provide further results and details in our online appendix.

Our methodology for RQ$_3$ is to follow the same procedures as in RQ$_1$ and RQ$_2$, except to compare models augmented with our whole-project encoder to models augmented with a file context encoder proposed by Haque~\emph{et al.}~\cite{haque2020improved}.

For RQ$_4$, we measure the size and training time for each of the models during training for RQ$_1$.  We report these resource costs alongside performance improvements for those models.


\subsection{Datasets}
\label{sec:data}

We use a Java dataset of 2.1m Java methods from 28k projects created by LeClair~\emph{et al.}~\cite{leclair2019recommendations} under strict quality guidelines.  These guidelines were tested for their effect on code summarization results, namely that the training and test sets are split by project, so that data from the test set does not leak into the test set by virtue of being in the same project.  We do not use datasets from other papers because they tend to be drawn from the same set of projects on online, open-source repositories (namely, Github), they tend to be smaller, and they are not vetted to the degree as this Java dataset.

We made one key change to the dataset in this paper when compared to previous papers: we improved the filter for code clones among the subroutines.  The original configuration in the datasets filtered code clones only by exact duplicates.  Since then, a study by Microsoft Research determined that this filter was insufficient for some ML tasks related to code, and recommended a new filtration procedure~\cite{allamanis2019adverse}.  We applied that filter to the dataset in this study.  The results we report for baselines in our experiments may have different values (usually lower values) than reported in the original papers for those baselines even for the same datasets.  The reason for this difference is our stricter removal of code clones.  It was necessary to rerun all experiments rather than rely on results reported in earlier papers, though the only difference in data or configuration was the code clone filtration procedure.

\subsection{Baselines}

We use four baseline neural code summarization techniques.  We then augment each with our whole-project encoder.  At a technical level, we build our implementations in a framework provided by Haque~\emph{et al.}~\cite{haque2020improved} in their reproducibility package for their paper on file context encoding.

{\small \textbf{\texttt{attendgru}}} is a typical seq2seq-like design like those used in early neural code summarization papers (and mentioned in Section~\ref{sec:impl}).  The only input to the encoder is the text from the source code of the subroutine itself.

{\small \textbf{\texttt{ast-attendgru}}} was proposed by LeClair~\emph{et al.}~\cite{leclair2019neural} and built on {\small \texttt{attendgru}} as well as work by Hu~\emph{et al.}~\cite{hu2018deep}.  It uses a flattened AST to represent subroutines.

{\small \textbf{\texttt{graph2seq}}} is a representative example in a recent class of graph neural network (GNN)-based techniques.  These techniques use information extracted from the AST and other relationships in code~\cite{leclair2020improved, liu2021retrievalaugmented, zugner2021languageagnostic}.

{\small \textbf{\texttt{code2seq}}} is a representative of AST path-based representations of code.  This baseline is a faithful reimplementation of a model proposed by Alon~\emph{et al.}~\cite{alon2018code2seq}, though with several hyperparameter changed to match those in other baselines.

We denote the versions of these baselines augmented with our whole-project encoder with the suffix {\small \textbf{\texttt{-pc}}} for ``project context.''  For example,  {\small \texttt{attendgru-pc}} and {\small \texttt{code2seq-pc}}.

For RQ$_3$, we use the file context encoder from Haque~\emph{et al.}~\cite{haque2020improved} as a baseline.  Models augmented with the file context encoder are denoted with the suffix {\small \textbf{\texttt{-fc}}}.

\subsection{Software / Hardware Details}
\label{sec:hardware}

Our hardware platform consisted of an HP Z-640 workstation with a Xeon E-1650v4 CPU, 128GB system memory, and two Nvidia Quadro P5000 GPUs with 16GB VRAM each.  Key software included CUDA 10.0 and Tensorflow 2.4.

\subsection{Threats to Validity}

The key threats to validity to this study include the datasets and the implementation details.  We chose a vetted dataset with millions of examples, but it is possible that results may not generalize to all datasets or other languages.  Likewise, results may vary given the plethora of implementation decisions, such as the means of combining the whole-project encoder output with other encoder output.  Caution is advised in drawing conclusions from these results beyond the scope of large open-source dataset in Java, or when implementation details differ significantly from those presented.

\section{Evaluation Results}

In this section, we discuss our evaluation results, including answers to our research questions and supporting analysis.

\newcommand\RotText[1]{\fontsize{9}{9}\selectfont
	\rotatebox[origin=c]{90}{\parbox{2.6cm}{%
			\centering#1}}}

\begin{figure*}[]
	\centering
	\small
\begin{tabular}{lllllllllll}
	&                                       & \multicolumn{5}{c}{BLEU}                                                                                                   & \multicolumn{3}{c}{ROUGE-LCS}                                         &  \\
	&  \multicolumn{1}{p{2.85cm}}{~}              & \multicolumn{1}{c}{A} & \multicolumn{1}{c}{1} & \multicolumn{1}{c}{2} & \multicolumn{1}{c}{3} & \multicolumn{1}{c|}{4}     & \multicolumn{1}{c}{P} & \multicolumn{1}{c}{R} & \multicolumn{1}{c}{F} & \multirow{10}{*}{\hspace{-0.1cm}\includegraphics[width=0.3\textwidth]{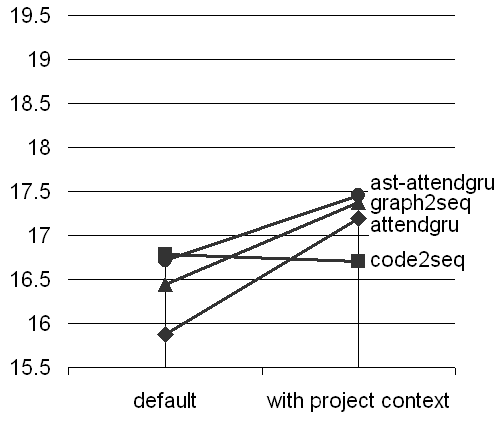}}                   \\
	\multirow{12}{*}{\RotText{\textbf{Solo Models}}}& \multicolumn{1}{l|}{attendgru}       & \textbf{15.87}        & 36.22                 & 18.89                 & 11.55                 & \multicolumn{1}{l|}{{\color{white}0}8.03} & 55.22                 & 46.98                 & 48.94                 &                    \\
	& \multicolumn{1}{l|}{attendgru-fc}      & \textbf{16.67}        & 36.53                 & 19.59                 & 12.31                 & \multicolumn{1}{l|}{{\color{white}0}8.76} & 56.54                 & 47.44                 & 49.78                 &                    \\
	& \multicolumn{1}{l|}{attendgru-pc}      & \textbf{17.19}        & 37.34                 & 20.20                 & 12.71                 & \multicolumn{1}{l|}{{\color{white}0}9.10} & 56.12                 & 47.81                 & 49.88                 &                    \\ \cline{3-10}
	& \multicolumn{1}{l|}{ast-attendgru}     & \textbf{16.72}        & 37.18                 & 19.84                 & 12.30                 & \multicolumn{1}{l|}{{\color{white}0}8.62} & 56.16                 & 47.83                 & 49.84                 &                    \\
	& \multicolumn{1}{l|}{ast-attendgru-fc}     & \textbf{17.18}        & 37.64                 & 20.21                 & 12.69                 & \multicolumn{1}{l|}{{\color{white}0}9.03} & 55.86                 & 48.11                 & 49.94                 &                    \\
	& \multicolumn{1}{l|}{ast-attendgru-pc}     & \textbf{17.45}        & 37.34                 & 20.47                 & 13.02                 & \multicolumn{1}{l|}{{\color{white}0}9.32} & 57.08                 & 48.05                 & 50.37                 &                    \\ \cline{3-10}
	& \multicolumn{1}{l|}{graph2seq}     & \textbf{16.44}        & 36.14                 & 19.54                 & 12.18                 & \multicolumn{1}{l|}{{\color{white}0}8.49} & 57.43                 & 47.32                 & 50.01                 &                    \\
	& \multicolumn{1}{l|}{graph2seq-fc}     & \textbf{16.26}        & 35.95                 & 19.19                 & 11.95                 & \multicolumn{1}{l|}{{\color{white}0}8.48} & 56.56                 & 46.95                 & 49.50                 &                    \\
	& \multicolumn{1}{l|}{graph2seq-pc}    & \textbf{17.37}        & 37.61                 & 20.37                 & 12.89                 & \multicolumn{1}{l|}{{\color{white}0}9.21} & 55.70                 & 47.87                 & 49.74                 &                    \\ \cline{3-10}
	& \multicolumn{1}{l|}{code2seq}     & \textbf{16.78}        & 37.88                 & 20.14                 & 12.30                 & \multicolumn{1}{l|}{{\color{white}0}8.45} & 55.45                 & 48.32                 & 49.80                 &                    \\
	& \multicolumn{1}{l|}{code2seq-fc}     & \textbf{16.45}        & 36.34                 & 19.43                 & 12.09                 & \multicolumn{1}{l|}{{\color{white}0}8.58} & 56.66                 &47.29                  & 49.79                 &                    \\
	& \multicolumn{1}{l|}{code2seq-pc}    & \textbf{16.7}        & 35.92                 & 19.53                 & 12.44                 & \multicolumn{1}{l|}{{\color{white}0}8.91} & 57.75                 & 47.08                 & 50.01                 &                                 
\end{tabular}

\vspace{0.4cm}

\begin{tabular}{llllllllllll}
	& &                                       & \multicolumn{5}{c}{BLEU}                                                                                                   & \multicolumn{3}{c}{ROUGE-LCS}                                         &  \\
	& & \multicolumn{1}{c}{mix}          & \multicolumn{1}{c}{A} & \multicolumn{1}{c}{1} & \multicolumn{1}{c}{2} & \multicolumn{1}{c}{3} & \multicolumn{1}{c|}{4}     & \multicolumn{1}{c}{P} & \multicolumn{1}{c}{R} & \multicolumn{1}{c}{F} & \multirow{10}{*}{\hspace{-0.1cm}\includegraphics[width=0.3\textwidth]{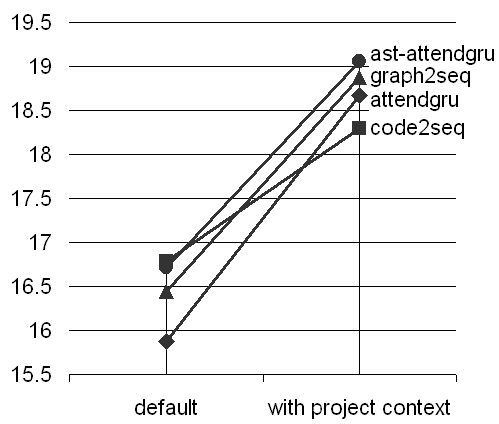}}                   \\
	\multirow{12}{*}{\RotText{\textbf{Ensemble Models}}}& attendgru & \multicolumn{1}{l|}{nc+fc}       & \textbf{17.96}        & 37.83                 & 20.92                 & 13.49                 & \multicolumn{1}{l|}{{\color{white}0}9.74} & 58.26                 & 48.79                 & 51.29                 &                    \\
	& attendgru    & \multicolumn{1}{p{0.75cm}|}{nc+pc} & \textbf{18.18}        & 38.26                 & 21.14                & 13.64                 & \multicolumn{1}{l|}{{\color{white}0}9.91} & 57.80                 & 48.98                 & 51.18                 &                    \\
	& attendgru    & \multicolumn{1}{l|}{fc+pc} & \textbf{18.67}        & 38.63                 & 21.62                 & 14.10                 & \multicolumn{1}{l|}{10.32} & 58.49                 & 49.29                 & 51.71                 &                    \\ \cline{3-11}
	& ast-attendgru    & \multicolumn{1}{l|}{nc+fc} & \textbf{18.47}        & 38.68                 & 21.53                 & 13.88                 & \multicolumn{1}{l|}{10.07} & 58.15                 & 49.33                 & 51.57                 &                    \\
	& ast-attendgru    & \multicolumn{1}{l|}{nc+pc} & \textbf{18.84}        & 38.80                 & 21.88                 & 14.26                 & \multicolumn{1}{l|}{10.41} & 58.71                 & 49.50                 & 51.91                 &                    \\
	& ast-attendgru    & \multicolumn{1}{l|}{fc+pc} & \textbf{19.06}        & 39.09                 & 22.07                 & 14.43                 & \multicolumn{1}{l|}{10.60} & 58.50                 & 49.63                 & 51.93                 &                    \\ \cline{3-11}
	& graph2seq    & \multicolumn{1}{l|}{nc+fc} & \textbf{17.77}        & 37.32                 & 20.79                 & 13.37                 & \multicolumn{1}{l|}{{\color{white}0}9.62} &59.16                  &48.58                  & 51.50                 &                    \\
	& graph2seq    & \multicolumn{1}{l|}{nc+pc} & \textbf{18.59}        & 38.85                 & 21.56                 & 14.06                 & \multicolumn{1}{l|}{10.28} & 58.54                 & 49.07                 & 51.58                 &                    \\
	& graph2seq   & \multicolumn{1}{l|}{fc+pc} & \textbf{18.80}        & 38.62                 & 21.72                 & 14.23                 & \multicolumn{1}{l|}{10.48} & 58.56                 & 49.20                 & 51.70                 &                    \\ \cline{3-11}
	& code2seq    & \multicolumn{1}{l|}{nc+fc} & \textbf{18.16}        & 38.35                 & 21.24                 & 13.60                 & \multicolumn{1}{l|}{{\color{white}0}9.81} &58.14                 & 49.14                 & 51.47                 &                    \\
	& code2seq    & \multicolumn{1}{l|}{nc+pc} & \textbf{18.56}        & 38.50                 & 21.55                 & 13.99                 & \multicolumn{1}{l|}{10.21} & 58.49                 & 49.23                 & 51.63                 &                    \\
	& code2seq   & \multicolumn{1}{l|}{fc+pc} & \textbf{18.29}        & 37.67                 & 21.12                 & 13.82                 & \multicolumn{1}{l|}{10.18} & 59.15                 & 48.72                 & 51.64                 &                                 
\end{tabular}

\vspace{0.1cm}
\caption{Results summary for RQ$_1$ and RQ$_3$.  The table shows the BLEU and ROUGE-LCS scores for baselines and our augmented versions of those baselines.  Chart depicts relative improvement of our augmented version according to aggregated BLEU score.  Column ``mix'' indicates which models were ensembled: nc for default/no-context, fc for file context, and pc for project context.}
\label{fig:rq1}
\vspace{-0.4cm}
\end{figure*}

\subsection{RQ$_1$: Effect of Augmenting Baselines}
\label{sec:rq1}

We found improved levels of BLEU and ROUGE scores across several baselines and configurations, when comparing default versions of the baselines to versions augmented with our project encoder.  Figure~\ref{fig:rq1} summarizes these results.  We report results under two key conditions: solo and ensemble.  A solo model is a single trained model -- it includes the model weights of the epoch which achieved the highest validation accuracy, under the training procedure described in Section~\ref{sec:metho}.  An ensemble model combines two trained models.  For example, the models for {\small \texttt{attendgru}} and {\small \texttt{attendgru-pc}} would be combined to form an ensemble model denoted ``nc+fc'' (no context plus file context, see first column of Ensemble Models table in Figure~\ref{fig:rq1}).  The combination procedure is to calculate the element-wise mean of the output predictions from each model, as recommended by Garmash~\emph{et al.}~\cite{garmash2016ensemble}.  We use this procedure in light of experimental findings for ensemble neural code summarization models by LeClair~\emph{et al.}~\cite{leclair2019neural}.

Two observations stand out for the solo models.  First, for the three baselines {\small \texttt{attendgru}}, {\small \texttt{ast-attendgru}}, and {\small \texttt{graph2seq}}, aggregate BLEU score improves between 4 and 8\% when our project encoder is added to the model.  The greatest improvement occurred for {\small \texttt{attendgru}}, which rose from 15.87 to 17.19 BLEU.  Note that this version is the one described in our Integration example in Section~\ref{sec:impl}.  It shows that even a relatively simple baseline can achieve competitive BLEU scores by adding project context ({\small \texttt{attendgru}} is just a vanilla seq2seq-like model with a single unidirectional GRU in the encoder and decoder).

Higher performance is observed for {\small \texttt{ast-attendgru}} and {\small \texttt{graph2seq}}, which is expected based on previous studies~\cite{haque2020improved}.  The higher performance is because both model designs use information from the AST of the subroutine. The {\small \texttt{graph2seq}} model uses a GNN, while {\small \texttt{ast-attendgru}} flattens the tree and uses it as input to an RNN.  Note that while both models see an improvement with project context, it is lower in relative terms than for {\small \texttt{attendgru}}.  We attribute this lower relative improvement to the increased amount of information that the model must learn in the same size vector space close to the output layer of the model.  Recall from Section~\ref{sec:impl} paragraph 3 that two vectors of size $e$ are generated: one for the original encoder and one for the project context encoder.  Concatenating them results in a vector of length 2x$e$.  To control model complexity, it is necessary to squash these vectors back to size $e$ with a dense network.  It is likely that information is lost.  This effect probably also explains the \emph{drop} in performance for {\small \texttt{code2seq}}.  That baseline is extremely complex, and the vector size of $e$ may be too confining.

The ensemble results are, in general, much higher.  The chart at the lower-right of Figure~\ref{fig:rq1} shows the default solo configuration to the ``fc+pc'' ensemble test condition.  All models demonstrate considerable improvement, between 9 and 17.5\%  The reason for this improvement proffered by Garmash~\emph{et al.}~\cite{garmash2016ensemble} is that different models contribute more to some predictions than others.  Mathematically, it means that the value of the argmax in the output vector of some models will be higher than others, because that model recognized a pattern closely-associated with that prediction.  Haque~\emph{et al.}~\cite{haque2020improved} pointed out that file context-based models contribute more to some subroutines than others.  Project context helps overall, but there are still subroutines for which other models are more useful.  The best performance is achieved by combining them.

\begin{figure}[!b]
	\small
	\centering
	\vspace{-0.6cm}
	\begin{tabular}{cc}
		\includegraphics[width=0.20\textwidth]{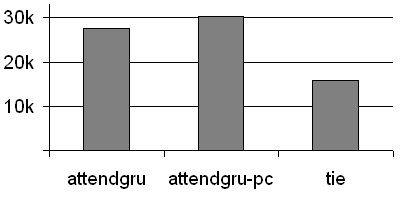}	\hspace{0.05cm} & \hspace{0.05cm} \includegraphics[width=0.20\textwidth]{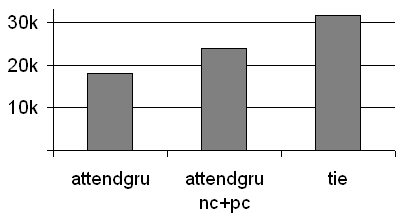} 
	\end{tabular}
	\vspace{-0.1cm}
	\caption{(left) Number of subroutines in the test set for which {\texttt{attendgru}} and {\texttt{attendgru-pc}} each had a higher BLEU score, and the number of ties. (right) Comparison of solo {\texttt{attendgru}} to ensemble {\texttt{attendgru nc+pc}}.}
	\label{fig:cmprq1}
\end{figure}

Consider Figure~\ref{fig:cmprq1}.  The improvement of different models is not necessarily distributed equally over all subroutines.  The chart on the left shows that of 73k subroutines in the test set, {\small \texttt{attendgru}} earned the highest score for about 27.5k, compared to about 30k for {\small \texttt{attendgru-pc}}.  This means that the reason {\small \texttt{attendgru-pc}} improved is because it created better predictions for only a portion of the results.  In that light, consider the chart on the right.  That chart compares solo {\small \texttt{attendgru}} to the ``nc+pc'' ensemble.  It shows that there is a substantial subset of subroutines for which {\small \texttt{attendgru}} earns a higher BLEU score, even compared to the ``nc+pc'' ensemble of {\small \texttt{attendgru}} and {\small \texttt{attendgru-pc}}.  What changes is that there is a much higher number of ties.  The ensemble sometimes creates predictions more like {\small \texttt{attendgru}}, and likewise more like {\small \texttt{attendgru-pc}} for other subroutines.  What is happening is that the output vector from {\small \texttt{attendgru}} has higher values for the predictions where it finds patterns closely associated with those predictions -- when it does not find those patterns, the values are lower and {\small \texttt{attendgru-pc}} is often higher.  The result is a better overall BLEU score.

\subsection{RQ$_2$: Action Word Prediction}

We found broadly similar performance in terms of precision and recall for action word prediction.  Recall that Haque~\emph{et al.}~\cite{haque2021action} recently recommended focusing on the prediction of the action word in source code summaries, given that word's importance in the summary.  Following these recommendations, we report the top-40, top-10, top-10n (which is the top 2-12, skipping get/set), and get/set.  The model should be able to distinguish get from set with very high accuracy, the top-10 and top-10n results being a more difficult problem, and the top-40 with even more difficulty.  The idea is that if the model cannot even predict the correct action word, then it may have little hope of predicting the rest of the summary.

We do not observe a large difference attributable to file or project context.  Our interpretation of this result is that the subroutine itself tends to provide most of the information needed to predict the action word -- many times the correct action word is in the name of the function e.g. ``book'' for {\small \texttt{book()}} in the class {\small \texttt{Seat}} in project {\small \texttt{AircraftTravel}}.  The higher BLEU and ROUGE scores for project context must therefore be due to improvements in the prediction of other parts of the code summary.  For example, if the summary is ``book seat on airplane'', the subroutine name will provide the action word ``book'', but code context will help find ``seat'' and ``airplane.''  We explore an example like this in Section~\ref{sec:descr}.

{\setlength\extrarowheight{5pt}\centering\setlength\tabcolsep{2pt}
	\begin{figure}[!t]
		\centering
		\scriptsize
		\vspace{0.1cm}
		\begin{tabular}{rlllllllllll}
			\multicolumn{1}{l}{\vspace{-0.5cm}\multirow{2}{*}{\textbf{Java}}} &        &     &     &     &        &            &      &        &       &    \\
			\multicolumn{1}{l}{top-10}                                 & \rotatebox{90}{return} & \rotatebox{90}{set} & \rotatebox{90}{get} & \rotatebox{90}{add} & \rotatebox{90}{create} & \rotatebox{90}{initialize} & \rotatebox{90}{test} & \rotatebox{90}{remove} & \rotatebox{90}{check} & \rotatebox{90}{is} &\rotatebox{90}{\emph{other}} \\
			return                                               &6160&32   &2365&10  &72  &14  &6   &6   &262&1 &2093\\
			set                                                  &11  &10388&13  &40  &9   &15  &2   &2   &1  &0 &558 \\
			get                                                  &4491&22   &3210&7   &13  &14  &2   &2   &30 &0 &907 \\
			add                                                  &4   &46   &1   &2950&14  &1   &10  &3   &1  &0 &293 \\
			create                                               &62  &21   &25  &43  &980 &16  &16  &0   &2  &0 &538 \\
			initialize                                           &23  &14   &3   &2   &22  &1632&0   &0   &0  &0 &201 \\
			test                                                 &35  &2    &2   &1   &2   &1   &1232&0   &61 &0 &170 \\
			remove                                               &6   &1    &0   &1   &0   &1   &3   &1186&0  &0 &303 \\
			check                                                &244 &9    &3   &2   &3   &0   &66  &3   &526&3 &607 \\
			is                                                   &95  &39   &10  &15  &13  &7   &4   &7   &49 &29 &366 \\
			\emph{other}                                                   &2715 &2147   &1023  &517  &605  &256   &727   &190   &378 &3 &22266 \\
		\end{tabular}
	\end{figure}
	\vspace{0.1cm}
}

{\setlength\tabcolsep{3pt}
	\begin{figure}[!t]
		\centering
		\scriptsize
		\begin{tabular}{lllllllllllll}
			\multirow{2}{*}{\textbf{Java}}      & \multicolumn{3}{c}{top-40}    & \multicolumn{3}{c}{top-10}    & \multicolumn{3}{c}{top-10n}   & \multicolumn{3}{c}{get/set} \\
			& ~p & ~r & ~f                     & ~p & ~r & ~f                     & ~p & ~r & ~f                     & ~p       & ~r       & ~f       \\
			\multicolumn{1}{l|}{attendgru}        &.53&.44& \multicolumn{1}{l|}{.45} &.69&.61& \multicolumn{1}{l|}{.60} &.70&.54& \multicolumn{1}{l|}{.54} &.99&.99&.99         \\
			\multicolumn{1}{l|}{attendgru-fc}        &.52&.46& \multicolumn{1}{l|}{.47} &.68&.63& \multicolumn{1}{l|}{.62} &.72&.51& \multicolumn{1}{l|}{.53} &.99&.99&.99         \\
			\multicolumn{1}{l|}{attendgru-pc}        &.53&.47& \multicolumn{1}{l|}{.47} &.65&.62& \multicolumn{1}{l|}{.61} &.71&.53& \multicolumn{1}{l|}{.54} &.99&.99&.99         \\
			\multicolumn{1}{l|}{ast-attendgru}    &.54&.46& \multicolumn{1}{l|}{.47} &.69&.61& \multicolumn{1}{l|}{.61} &.72&.52& \multicolumn{1}{l|}{.53} &.99&.99&.99         \\
			\multicolumn{1}{l|}{ast-attendgru-fc} &.55&.47& \multicolumn{1}{l|}{.47} &.61&.61& \multicolumn{1}{l|}{.60} &.72&.51& \multicolumn{1}{l|}{.53} &.99&.99&.99         \\
			\multicolumn{1}{l|}{ast-attendgru-pc} &.55&.44& \multicolumn{1}{l|}{.46} &.67&.63& \multicolumn{1}{l|}{.62} &.68&.52& \multicolumn{1}{l|}{.53} &.99&.99&.99         \\
			\multicolumn{1}{l|}{graph2seq}        &.56&.45& \multicolumn{1}{l|}{.46} &.68&.61& \multicolumn{1}{l|}{.61} &.69&.54& \multicolumn{1}{l|}{.55} &.99&.99&.99         \\
			\multicolumn{1}{l|}{graph2seq-fc}        &.55&.45& \multicolumn{1}{l|}{.46} &.70&.60& \multicolumn{1}{l|}{.61} &.68&.52& \multicolumn{1}{l|}{.54} &.99&.99&.99         \\
			\multicolumn{1}{l|}{graph2seq-pc}        &.56&.45& \multicolumn{1}{l|}{.46} &.70&.59& \multicolumn{1}{l|}{.60} &.71&.51& \multicolumn{1}{l|}{.54} &.99&.99&.99         \\
			\multicolumn{1}{l|}{code2seq}        &.54&.47& \multicolumn{1}{l|}{.46} &.70&.59& \multicolumn{1}{l|}{.59} &.71&.49& \multicolumn{1}{l|}{.52} &.99&.99&.99         \\
			\multicolumn{1}{l|}{code2seq-fc}        &.53&.47& \multicolumn{1}{l|}{.47} &.71&.59& \multicolumn{1}{l|}{.60} &.71&.53& \multicolumn{1}{l|}{.54} &.99&.99&.99     	  \\
			\multicolumn{1}{l|}{code2seq-pc}        &.54&.46& \multicolumn{1}{l|}{.47} &.68&.62& \multicolumn{1}{l|}{.62} &.69&.53& \multicolumn{1}{l|}{.54} &.99&.99&.99     
		\end{tabular}
		\vspace{0.3cm}
		\caption{\emph{(top)} Confusion matrix showing results for top-10 action words for ast-attendgru-pc, the best performer in terms of f-measure.  \emph{(bottom)} Overall results under standard conditions in the Java dataset.}
		\label{tab:rq2}
		\vspace{-0.25cm}
	\end{figure}
}

\subsection{RQ$_3$: Comparison to File Context}


\begin{figure}[!b]
	\small
	\centering
	\vspace{-0.35cm}
	\begin{tabular}{c}
		\includegraphics[width=0.30\textwidth]{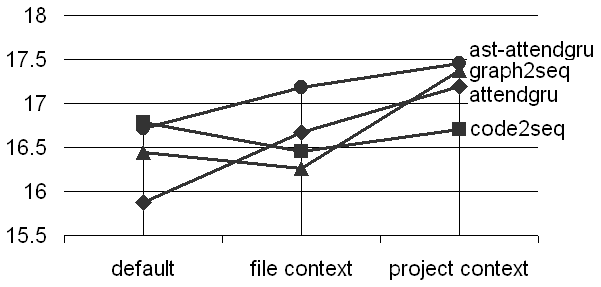}	\\
		(a) Solo \vspace{-0.15cm} \\
		 \\
		\includegraphics[width=0.30\textwidth]{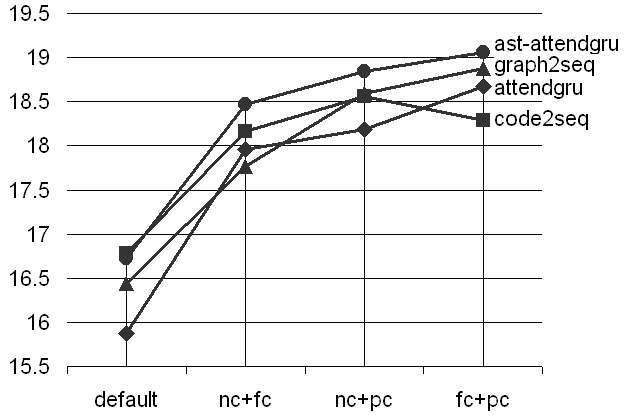}  \\
		(b) Ensemble
	\end{tabular}
	\caption{Comparison of aggregate BLEU scores for {\texttt{-fc}} and {\texttt{-pc}} models.  This figure is a depiction of values in Figure~\ref{fig:rq1}.}
	\label{fig:rq3}
\end{figure}

We observe improvement over the baselines that are enhanced with file context.  Figure~\ref{fig:rq3} depicts the change in aggregate BLEU score across key model configurations.  Figure~\ref{fig:rq3}a shows the default baseline model, followed by the file context and project context versions of those models.  Figure~\ref{fig:rq3}b shows the default baseline model, followed by the nc+fc and nc+pc ensembles.  Recall that the ``file context'' versions are those provided by Haque~\emph{et al.}~\cite{haque2020improved} and form the nearest competition for models that include code context (see Section~\ref{sec:related_scs}).

Overall, the project context versions of the baselines achieve higher aggregate BLEU scores than the file context versions.  However, the gains are not uniform.  For example, note in Figure~\ref{fig:rq3}a that {\small \texttt{graph2seq-fc}} is the lowest performing file context model, while {\small \texttt{graph2seq-pc}} is nearly tied for the top position of solo models.  This finding seems to be at odds with scores reported by Haque~\emph{et al.}~\cite{haque2020improved} for {\small \texttt{graph2seq-fc}}.  We attribute the difference to the enhanced removal of code clones we performed for experiments in this paper (see Section~\ref{sec:data}).  The file context contains many subroutines that are considered clones by the recommend clone removal technique~\cite{allamanis2019adverse}, because these subroutines may be overloaded or only slightly modified.  Future researchers using file context may consider leaving clones in the file context, and only remove them from the list of subroutines in the test set to ensure fairness.

The ensemble models also show that project context helps achieve higher BLEU scores than file context.  Figure~\ref{fig:rq3}b shows marked improvement from nc+fc ensembles (no context combined with file context) to nc+pc ensembles, then again from nc+pc to fc+pc ensembles.  The exception is {\small \texttt{code2seq}}, which we believe is due to the same vector size restriction described in Section~\ref{sec:rq1}.  The gain of fc+pc over other ensembles implies that file context and project context contributes to model predictions in orthogonal ways, in the same vein as the ensemble results for RQ$_1$, above.

\subsection{RQ$_4$: Effects on Model Size}

Adding project context to a baseline increases the complexity of the model, and this complexity comes at a cost in terms of time to train.  While it may be tempting to write off training time as a ``one time sunk cost'', in fact this added time imposes engineering challenges that affect cost-benefit decisions~\cite{hazelwood2018applied}.  We report training time per epoch as a proxy for this complexity cost.  All data points were collected on the hardware platform described in Section~\ref{sec:hardware}.

\begin{figure}[!h]
	\small
	\centering
	\vspace{-0.2cm}
	\includegraphics[width=0.33\textwidth]{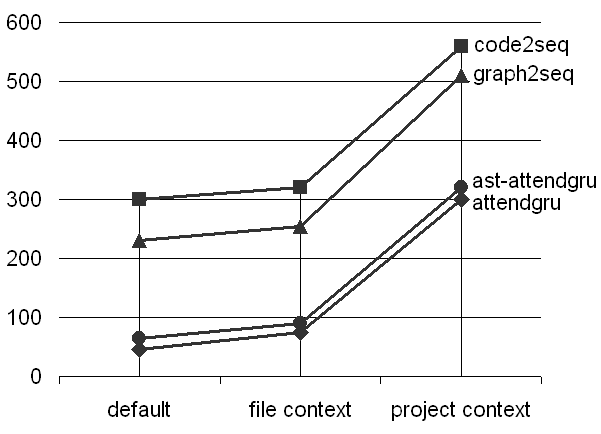}
	\vspace{-0.3cm}
	\caption{Training time in minutes per epoch.}
	\vspace{-0.2cm}
\end{figure}

We observe about a 3x time cost for {\small \texttt{attendgru}} and {\small \texttt{ast-attendgru}}, and about a 2x cost for {\small \texttt{graph2seq}} and {\small \texttt{code2seq}}, when comapring default configurations to the versions with project context.  While the number of minutes is subject to hardware and software settings, we report these numbers to assist practitioners in deciding how to deploy these technologies.  For instance, time required for {\small \texttt{ast-attendgru-pc}} is roughly equal to {\small \texttt{code2seq-fc}} even though {\small \texttt{ast-attendgru-pc}} achieves a higher BLEU score. At that time limit {\small \texttt{ast-attendgru-pc}} may be the best choice even if {\small \texttt{code2seq}} is a better baseline than {\small \texttt{ast-attendgru}}.

An implication of this finding is that the costs of including project context can be very high.  We find improvement in terms of overall BLEU score, especially for ensemble models, the training difficulty is 2x to 3x even for a modest setting of $f$=10.  Future researchers may note the potential of even a portion of project context for improving prediction performance, and may consider guiding effort into reducing the costs so that more context may be considered.

\section{Discussion \& Conclusion}
\label{sec:descr}

This paper advances the state of the art in two ways:

\begin{enumerate}
	\item We propose an encoder that creates a vectorized representation of project context for use in neural models of software source code.
	\item We demonstrate the benefit of this encoder for the specific problem of source code summarization.
\end{enumerate}

The first advancement is important because of its potential impact on many areas of software engineering research.  Allamanis~\emph{et al.}~\cite{allamanis2018survey} present a survey of neural models for various software engineering tasks, and separate these tasks into two categories: code generational and code representational.  A code generational task is like code completion or automatic repair, in which the model is expected to create new source code.  A code representation task is like code summarization or bug localization, in which the model must create some internal representation of the program, and use it to predict something about the software, such as a code summary or if a subroutine contains a particular kind of bug.

The project context encoder we propose has potential in many code representational tasks.  Essentially what the encoder does is create a vectorized representation of the files surrounding a particular area of code in a project.  This representation could be used in many ways.  For example, a neural model for predicting bugs based on the code in a subroutine could append our project context encoder.  It may help the model learn that a particular pattern in the code may be associated with bugs for some types of projects, but not others.  This benefit is only a hypothetical discussion -- the point is that this paper may have benefits beyond the specific problem of code summarization.

The implementation and experiment in this paper focus on code summarization, and demonstrate how project context improves predictions in terms of BLEU and ROUGE scores for several baselines (RQ$_1$).  We show that our model seems to be providing orthogonal information by improving predictions for a subset of subroutines, and that by using an ensemble procedure, the benefits of project context can be combined with file context and default models.  We also show that these improvements seem to be focused on areas outside the action word (RQ$_2$), and that project context tends to result in overall better scores that file context alone (RQ$_3$), even if an ensemble has the highest observed performance (fc+pc ensemble models, RQ$_1$).  Our project context encoder leads to an advancement of the state of the art for code summarization.

The reason that project context helps is that many methods are very difficult to understand from only the source code of a single subroutine.  Code summarization techniques that consider only the subroutine itself make the tacit assumption that that subroutine contains all the information necessary to summarize it.  Consider Example~\ref{fig:ex1} (method \#29987000 in the dataset~\cite{leclair2019recommendations}, we list ID numbers for reproducibility).  This method is from a GUI program for managing config files.  Its purpose is to stop and cleanup a plugin.  However, this purpose is hard to ascertain without seeing the project context.

Compare the predicted summary for Example~\ref{fig:ex1} by {\small \texttt{attendgru}} versus {\small \texttt{attendgru-pc}}.  The model {\small \texttt{attendgru}} predicts ``stops the bundle'', which seems a reasonable guess considering that it has access only to the source code of that subroutine.  The method is called {\small \texttt{stop()}}, which begs the question ``stops what?''  The word ``plugin'' is in the source code, but so is the word ``translator'', ``messages'', ``bundle'', etc.  Many methods in the training set have a pattern in which the action word is followed by a word from the parameter list~\cite{haque2021action}, and a simple seq2seq-like model such as {\small \texttt{attendgru}} can learn this pattern effectively~\cite{leclair2019neural, hu2018deep}.  So {\small \texttt{attendgru}} guesses ``stops the bundle.''

\begin{figure}[t!]
	\setcounter{figure}{0}
	\renewcommand{\figurename}{Example}
	
	{\small	
	\textbf{Method \#29987000:} (from test set)
	\vspace{0.2cm}

	\begin{tabular}{lm{6.4cm}} 
		\emph{reference} & this method is called when the plug in is stopped  	\\ \hline
		attendgru        & stops the bundle   	\\ 
		attendgru-pc     & this method is called when the plug in is stopped 		\\ 
		
	\end{tabular}
	}

	{\small
		
		\begin{verbatim}
  public void stop(BundleContext context)
      throws Exception {
    super.stop(context);
    plugin = null;
    Translator.removeAllMessages();
    Translator.removeAllTranslatables();
  }
		\end{verbatim}
	}

	{\small

		\vspace{0.0cm}
		
		\begin{tabular}{ll}
			\multicolumn{2}{l}{files in net.confex.application:}  \\
			~  & ApplicationWorkbenchAdvisor.java    \\
			~  & ApplicationWorkbenchWindowAdvisor.java         \\
			~  & ConfexApplication.java \\
			$\rightarrow$  & ConfexPlugin.java                             \\
			~  & Perspective.java \\
			~  & ToolbarLayout.java     \\                                         
		\end{tabular}
	}
\caption{A method from the test set for which attendgru-pc wrote the correct summary, while attendgru did not.  Method is in the project {net.confex.application}, which contains the six listed files.}
\label{fig:ex1}
\vspace{-0.1cm}
\end{figure}

\begin{figure}[t!]
	{
		\renewcommand{\figurename}{Example}

	{\small	
	\textbf{Method \#805539:} (from training set)
	\vspace{0.1cm}		

	\begin{tabular}{lm{6.4cm}} 
		\emph{reference} & this method is called when the plug in is stopped  	\\
		
	\end{tabular}
}

		{\small
			\begin{verbatim}
  public void stop(BundleContext context)
      throws Exception {
    super.stop(context);
    if (this.logManager != null) {
      this.logManager.shutdown();
      this.logManager = null;
    }
    if (searchProviderManager != null) {
      searchProviderManager.dispose();
      searchProviderManager = null;
    }
    plugin = null;
  }
	\end{verbatim}

	\begin{tabular}{ll}
	\multicolumn{2}{l}{files in net.bioclipse:}  \\
	~  & ApplicationWorkbenchAdvisor.java    \\
	~  & ApplicationWorkbenchWindowAdvisor.java         \\
	~  & ApplicationWorkbenchActionBarAdvisor.java         \\
	~ & BioclipsePerspective.java                             \\
	$\rightarrow$ & BioclipsePlugin.java \\
	~  & PerspectiveOpenPreferencePage.java     \\                                         
	\end{tabular}

		\caption{Method in the training set seen by both approaches.  Note list of files is similar to Example~\ref{fig:ex1} because both are built with the same GUI platform.  Project context helps attendgru-pc detect this similarity.}
		\label{fig:ex2}
		
		\vspace{-0.4cm}
	}
}
\end{figure}

The project context helps the model learn what the method really means.  The method in Example~\ref{fig:ex1} appears in the file {\small \texttt{ConfexPlugin.java}}, which is in the project with five other files.  While there is far too much data in these files to reprint here, one may note the similarity between these files and the files in Example~\ref{fig:ex2} (method ID numbers may be used to recover these files in the dataset).  Example~\ref{fig:ex2} is from the training set.  The content of the method itself is quite different from Example~\ref{fig:ex1} (aside from the method signature), so {\small \texttt{attendgru}} has difficulty seeing the two methods as similar -- there are many methods named {\small \texttt{stop()}} that have nothing to do with plugins.  But {\small \texttt{attendgru-pc}} has access to the project context and can identify the similarity of the files in this context.  As a result, {\small \texttt{attendgru-pc}} predicts the summary that it has learned during training, which is correct.

We caution that we selected these examples as a demonstration of what project context can offer, and may not be representative of how the model always behaves.  The model can make incorrect predictions -- recall from Section~\ref{sec:rq1} that there is a subset of methods for which project context outperforms the baseline, and a subset where it does not.  However, when the project context models go astray, it tends to be because they are recognizing patterns in the context, and we found that ensemble models can generate better summaries.

Our intent in this paper is to propose a technique for encoding the project context of source code.  While project context is quite expansive, our technique can capture enough of this context to be useful for the task of source code summarization.  Essentially what have shown is that even a small amount of this project context -- just $f$=10 in this paper -- can lead to significant improvements in the aggregate BLEU scores of several baselines.  In ensemble models, the benefit increases further.  However, as we observe in RQ$_5$, the costs of including even limited project context mushroom rapidly.  Future work aims to capture more of this context and interactions among the code components such as dependency relationships, and to demonstrate the benefit of context to other areas using neural models of source code.



\section{Reproducibility}
\label{sec:repro}

We strongly encourage reproducibility.  We provide the following online appendix to facilitate reuse of this technology by practitioners and other researchers.  Our code, dataset scripts, and operating instructions may be found at:

\texttt{https://github.com/aakashba/projcon}

\section*{Acknowledgment}
This work is supported in part by the NSF CCF-1452959 and CCF-1717607 grants. Any opinions, findings, and conclusions expressed herein are the authors’ and do not necessarily reflect those of the sponsors.

\bibliographystyle{IEEEtran}
\bibliography{main}

\end{document}